\documentclass[sigconf]{acmart}
\usepackage{algorithm}
\usepackage{algpseudocode}
\usepackage{tikz}
\usepackage{pgfplots}

\usepackage{booktabs} % For formal tables

\providecommand{\tightlist}{%
  \setlength{\itemsep}{0pt}\setlength{\parskip}{0pt}}

\setcopyright{rightsretained}
\acmConference[KDD'2017]{Knowledge Discovery and Data Mining}{August 2017}{Halifax -- Canada}

\begin{document}

\title{A linear streaming algorithm for community detection in very large
networks}

\author{Alexandre Hollocou}
\affiliation{%
  \institution{INRIA}
  \city{Paris} 
  \state{France} 
}
\email{alexandre.hollocou@inria.fr}

\author{Julien Maudet}
\affiliation{%
  \institution{Ecole Polytechnique}
  \city{Palaiseau} 
  \state{France} 
}
\email{julien.maudet@polytechnique.edu}

\author{Thomas Bonald}
\affiliation{%
  \institution{Telecom Paristech}
  \city{Paris} 
  \state{France} 
}
\email{thomas.bonald@telecom-paristech.fr}

\author{Marc Lelarge}
\affiliation{%
  \institution{INRIA}
  \city{Paris} 
  \state{France} 
}
\email{marc.lelarge@ens.fr}

\begin{abstract}
In this paper, we introduce a novel community detection algorithm in
graphs, called SCoDA (Streaming Community Detection Algorithm), based on
an edge streaming setting. This algorithm has an extremely low memory
footprint and a lightning-fast execution time as it only stores two
integers per node and processes each edge strictly once. The approach is
based on the following simple observation: if we pick an edge uniformly
at random in the network, this edge is more likely to connect two nodes
of the same community than two nodes of distinct communities. We exploit
this idea to build communities by local changes at each edge arrival.
Using theoretical arguments, we relate the ability of SCoDA to detect
communities to usual quality metrics of these communities like the
conductance. Experimental results performed on massive real-life
networks ranging from one million to more than one billion edges shows
that SCoDA runs more than ten times faster than existing algorithms and
leads to similar or better detection scores on the largest graphs.
\end{abstract}

\date{}


\maketitle

\section{Introduction}\label{introduction}

\label{section-intro}

\subsection{Motivations}\label{motivations}

Networks arise in a wide range of fields from biology
\citep{palla2005uncovering} to social media
\citep{mislove2007measurement} or web analysis
\citep{flake2000efficient}\citep{pastor2007evolution}. In most of these
networks, we observe groups of nodes that are densely connected between
each other and sparsely connected to the rest of the graph. One of the
most fundamental problems in the study of such networks consists in
identifying these dense clusters of nodes. This problem is commonly
referred to as \emph{community detection}.

A major challenge for community detection algorithms is their ability to
process very large networks that are commonly observed in numerous
fields. For instance, social networks have typically millions of nodes
and billions of edges (e.g. Friendster \citep{mislove2007measurement}).
Many algorithms have been proposed during the last ten years, using
various techniques ranging from combinatorial optimization to spectral
analysis \citep{lancichinetti2009community}. Most of them fail to scale
to such large real-life networks \citep{prat2014high}.

\subsection{Contributions}\label{contributions}

In this paper, we introduce a novel approach to detect communities in
very large graphs. This approach is based on \emph{edge streams} where
network edges are streamed in a random order. The algorithm processes
each edge strictly once. Moreover, the algorithm only stores two
integers for each node: its current community index and the number of
adjacent edges that have already been processed. Hence, the time
complexity of the algorithm is linear in the number of edges and its
space complexity is linear in the number of nodes. In the experimental
evaluation of the algorithm we show that this streaming algorithm,
called SCoDA (Streaming Community Detection Algorithm), is able to
handle massive graphs \citep{yang2015defining} with low execution time
and memory consumption.

\subsection{Related work}\label{related-work}

A number of algorithms have been developed for detecting communities in
networks or graphs\footnote{In the rest of the paper,
we use the terms \textit{networks} and \textit{graphs}
interchangeably.} \citep{fortunato2010community}. Many rely on the
optimization of some objective function that measures the quality of the
detected communities. The most popular quality metric is the
\emph{modularity} \citep{newman2006modularity}, which is based on the
comparison between the number of edges that are observed in each cluster
and the number of edges that would be observed if the edges were
randomly distributed. Other metrics have been used with success, like
the conductance, the out-degree fraction and the clustering coefficient
\citep{yang2015defining}. Another popular class of algorithms uses
random walks \citep{pons2005computing}\citep{whang2013overlapping}.
These methods are based on the fact that random walks tend to get
``trapped'' in the dense zones of the graph. These techniques have
proved to be efficient but are often time-consuming and fail to scale to
large graphs \citep{prat2014high}. Other popular methods include
spectral clustering \citep{spielman2007spectral}\citep{von2007tutorial},
clique percolation \citep{palla2005uncovering}, statistical inference
\citep{lancichinetti2010statistical}, or matrix factorization
\citep{yang2013overlapping}.

The streaming approach has drawn considerable interest in network
analysis over the last decade. Within the data stream model, massive
graphs with potentially billions of edges can be processed without being
stored in memory \citep{mcgregor2014graph}. A lot of algorithm have been
proposed for different problems that arise in large networks, such as
counting subgraphs \citep{bar2002reductions}\citep{buriol2006counting},
computing matchings
\citep{goel2012communication}\citep{feigenbaum2005graph}, finding the
minimum spanning tree \citep{elkin2006efficient}\citep{tarjan1983data}
or graph sparsification \citep{benczur1996approximating}. Different
types of data streams can be considered: \emph{insert-only streams},
where the stream is the unordered sequence of the network edges, or
\emph{dynamic graph streams}, where edges can both be added or deleted.
Many streaming algorithms rely on \emph{graph sketches} which store the
input in a memory-efficient way and are updated at each step
\citep{ahn2012graph}.

In this paper, we use the streaming setting to define a novel community
detection algorithm. We use randomized insert-only edge streams and
define a minimal sketch, by storing only two integers per node.

\subsection{Paper outline}\label{paper-outline}

The rest of the paper is organized as follows. We first describe our
streaming algorithm, SCoDA, in Section \ref{section-algorithm}. In
Section \ref{section-experiments}, we evaluate experimentally the
performance of SCoDA on real-life networks and compare it to
state-of-the-art algorithms. A theoretical analysis of SCoDA is
presented in Section \ref{section-analysis}. The choice of the only
parameter of SCoDA, as used in the experiments, is justified in Section
\ref{section-parameter}. Section \ref{section-conclusion} concludes the
paper.

\section{A streaming algorithm for community
detection}\label{a-streaming-algorithm-for-community-detection}

\label{section-algorithm}

In this section, we define SCoDA, a streaming algorithm for community
detection in graphs.

\subsection{Notations}\label{notations}

We are given an unweighted and undirected graph \(G(V,E)\) where \(V\)
is the set of vertices and \(E\) the set of edges. We use \(n\) to
denote the number of nodes and \(m\) the number of edges. Without loss
of generality, we consider that \(V = \{1,...,n\}\).

Let \(A\) and \(B\) be two subsets of \(V\). We use the following
notations: \[\begin{split}
%e(u) & = \{ (v, w) \in E: v=u \text{ or } w=u\}\\
e(A) & = \{ (u,v) \in E : u \in A \text{ or } v \in A \} \\
e(A,B) & = \{ (u,v) \in E : u \in A \text{ and } v \in B \} \\
d_{A}(u) &= |\{ v \in A : (u,v) \in E \}| \\
\partial A &= \{ u \in A : \exists v \in V \setminus A, (u,v) \in E \} \\
\overline{A} &= V \setminus A \\
\end{split}\]

Note that we have \(e(A,A) \subset e(A)\).

\subsection{Motivation}\label{motivation}

Although there is no universal definition of what a community is, most
existing algorithms rely on the principle that nodes tend to be more
connected within a community than across communities. Hence, if we pick
uniformly at random an edge \(e\) in \(E\), this edge is more likely to
link nodes of the same community (i.e., \(e\) is an
\emph{intra-community} edge), than nodes from distinct communities
(i.e., \(e\) is an \emph{inter-community} edge). Equivalently, if the
edges of \(E\) are processed in a random order, we expect many
intra-community edges to arrive before the inter-community edges.

More formally, let \(C \subset V\) be a community that we want to
detect. If the edges of \(E\) are randomly drawn without replacement, we
can consider the event where the first \(k\) edges drawn in \(e(C)\) are
\emph{intra-community} edges, i.e. in \(e(C,C)\): \[\begin{split}
\text{Intra}_k(C) = & \text{the first $k$ edges that are drawn} \\
& \text{from $e(C)$ are in $e(C,C)$}
\end{split}\] The probability of this event is:
\[\mathbb{P}[\text{Intra}_k(C)] = \prod_{l=0}^{k-1} \frac{|e(C,C)| - l}{|e(C)| - l}
= \prod_{l=0}^{k-1}  (1 - \phi_l(C)),\] where
\[\phi_l(C) = \frac{|e(C,\overline{C})|}{|e(C,C)| +
|e(C,\overline{C})| - l},\] for all \(l=0,\ldots,k-1\). Observe that the
definition of \(\phi_0(C)\) is very close to that of the conductance
\(\varphi(C)\) of \(C\),
\[\varphi(C) = \frac{|e(C,\overline{C})|}{2 |e(C,C)| + |e(C,\overline{C})|}.\]
In particular,
\(\phi_0(C) = \frac{2\varphi(C)}{1 + \varphi(C)}\approx 2 \varphi(C)\)
for small values of the conductance. We refer to \(\phi_0(C)\) as the
\emph{pseudo-conductance} in the rest of the paper. It is well known
that \emph{good} communities are subsets of \(V\) with low conductance
\citep{shi2000normalized}. We then expect \(\phi_{l}(C)\) to be low for
small values of \(l\) if \(C\) is a good community and the probability
of picking an \emph{inter-community} edge within the first \(k\) edges
picked at random in \(e(C)\) to be low for small values of \(k\).

\subsection{A streaming approach}\label{a-streaming-approach}

This observation is used to design an algorithm that streams the edges
of the network in a random order. For each arriving edge \((u, v)\), the
algorithm places \(u\) and \(v\) in the same community if the edge
arrives \emph{early} (intra-community edge) and splits the nodes in
distinct communities otherwise (inter-community edge). In this
formulation, the notion of an \emph{early} edge is of course critical.
In the proposed algorithm, we consider that an edge \((u, v)\) arrives
\emph{early} if the current degrees of nodes \(u\) and \(v\), accounting
for previously arrived edges only, is low.

More formally, the first step of the algorithm consists in shuffling the
list of edges \(E\), i.e., in generating a random permutation of the
list of edges. The algorithm then considers edges in this particular
order, say \(e_1,e_2,\ldots,e_m\). Each node is initally in its own
community. For each new edge \(e_j = (u, v)\), the algorithm performs
one of the following actions:

\begin{itemize}
\tightlist
\item
  \(u\) joins the community of \(v\);
\item
  \(v\) joins the community of \(u\);
\item
  no action.
\end{itemize}

The choice of the action depends on the updated degrees \(d(u)\) and
\(d(v)\) of nodes \(u\) and \(v\), i.e., the degree computed using the
edges \(e_{1},...,e_{j}\). If \(d(u)\) or \(d(v)\) is greater than a
given threshold \(D\), then we do nothing; otherwise, the node with the
lowest degree joins the community of the other node.

\subsection{Algorithm}\label{algorithm}

The algorithm SCoDA is defined in Algorithm \ref{scoda}. It takes the
list of edges of the graph and one integer parameter \(D \ge 1\). The
algorithm builds two arrays \(d\) and \(c\) of size \(n\). At the end of
the algorithm, \(d_i\) is the degree of node \(i\), and \(c_i\) the
community of node \(i\). When the algorithm starts, each node has degree
zero and is in its own community (\(d_i=0\) and \(c_i=i\) for all
\(i\)). Then, the list of edges is shuffled and the main loop iterates
over the edges in this random order. For each new edge \(e_j=(u, v)\),
the degrees of \(u\) and \(v\) are updated. Then, if these degrees are
both lower than the threshold parameter \(D\), the node with the lower
degree joins the community of the other node. Otherwise, the communities
remain unchanged.

\begin{algorithm}
\caption{SCoDA}
\label{scoda}
\begin{algorithmic}[1]
\Require
List of edges $E$ between nodes $\{1,...,n\}$ and 
parameter $D \ge 1$
\State For all $i=1,...,n$, $d_i \leftarrow 0$ and $c_i \leftarrow i$
\State Shuffle the list of edges $E$
\For{$j = 1,...,|E|$}
    \State $(u, v) \leftarrow$ $j^{th}$ edge of $E$
    \State
    $d_{u} \leftarrow d_{u} + 1$
    and $d_{v} \leftarrow d_{v} + 1$
    \If{$d_{u} \leq D$ and $d_{v} \leq D$}
       \If{$d_{u} \leq d_{v}$} $c_{u} \leftarrow c_{v}$
       \Else\ $c_{v} \leftarrow c_{u}$ \EndIf
    \EndIf
\EndFor
\State \Return $(c_i)_{i=1,...,n}$
\end{algorithmic}
\end{algorithm}

Observe that, in case of equality \(d_u=d_v\le D\), \(v\) joins the
community of \(u\). Of course, this choice is arbitrary and can be made
random (e.g., \(u\) joins the community of \(v\) with probability
\(1/2\) and \(v\) joins the community of \(u\) with probability
\(1/2\)). Equivalently, the random shuffling of the list of edges may
include for each edge \(e=(u,v)\) a random choice between \((u,v)\) and
\((v,u)\).

\begin{figure}
\centering
\includegraphics[height=11cm]{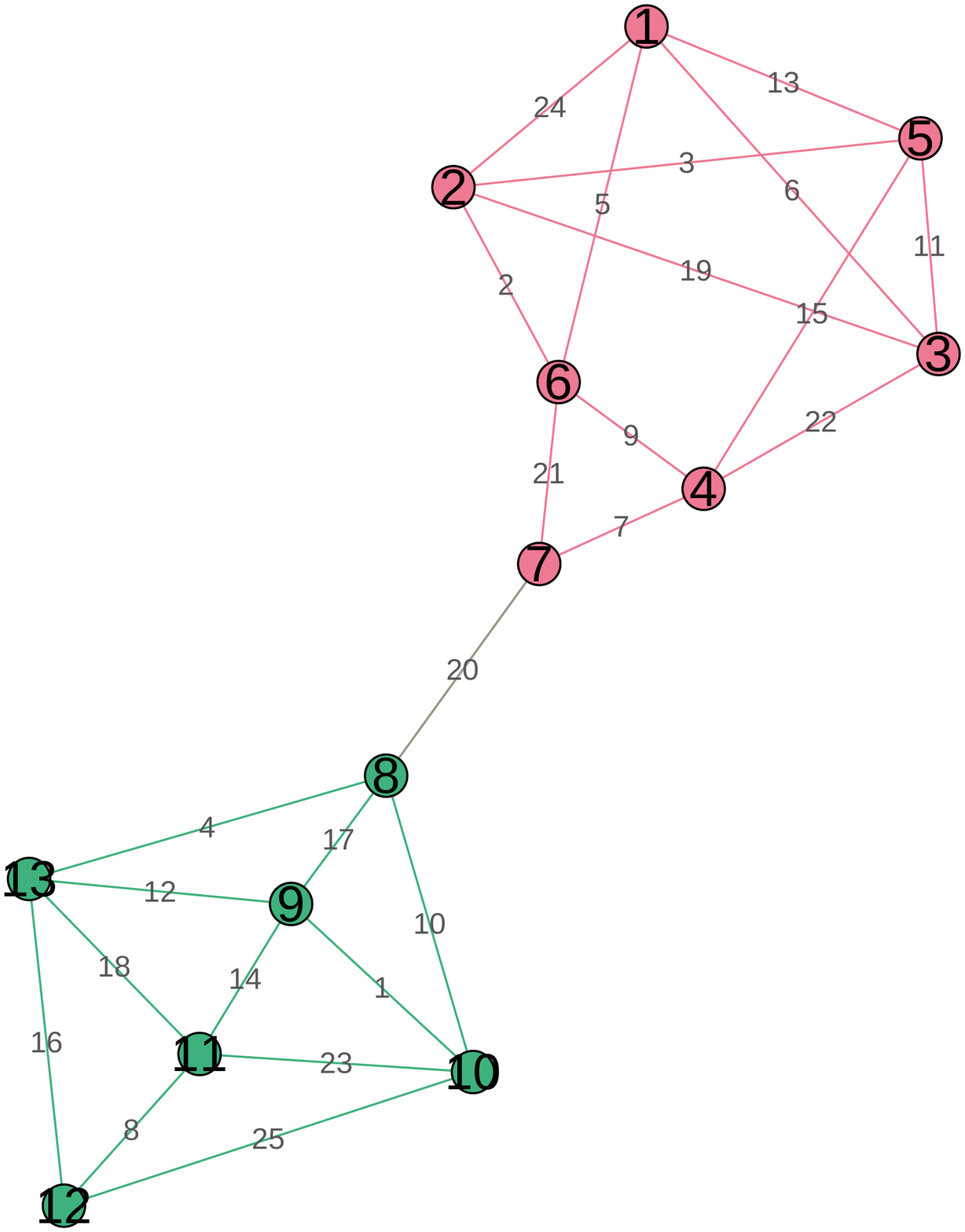}
\qquad
\resizebox{8.5cm}{!}{
\begin{tabular}{|l|l|}
\hline \textbf{Step 1} 
& $ C_{9} = [9, 10]$\\
\hline \textbf{Step 2} 
& $ C_{6} = [2, 6]$, $ C_{9} = [9, 10]$\\
\hline \textbf{Step 3} 
& $ C_{6} = [2, 5, 6]$, $ C_{9} = [9, 10]$\\
\hline \textbf{Step 4} 
& $ C_{6} = [2, 5, 6]$, $ C_{8} = [8, 13]$, $ C_{9} = [9, 10]$\\
\hline \textbf{Step 5} 
& $ C_{6} = [1, 2, 5, 6]$, $ C_{8} = [8, 13]$, $ C_{9} = [9, 10]$\\
\hline \textbf{Step 6} 
& $ C_{6} = [1, 2, 3, 5, 6]$, $ C_{8} = [8, 13]$, $ C_{9} = [9, 10]$\\
\hline \textbf{Step 7} 
& $ C_{6} = [1, 2, 3, 5, 6]$, $ C_{7} = [4, 7]$, $ C_{8} = [8, 13]$, \\
& $ C_{9} = [9, 10]$\\
\hline \textbf{Step 8} 
& $ C_{12} = [11, 12]$, $ C_{6} = [1, 2, 3, 5, 6]$, $ C_{7} = [4, 7]$, \\
& $ C_{8} = [8, 13]$, $ C_{9} = [9, 10]$\\
\hline \textbf{Step 9} 
& $ C_{12} = [11, 12]$, $ C_{6} = [1, 2, 3, 4, 5, 6]$, $ C_{8} = [8, 13]$, \\
& $ C_{9} = [9, 10]$\\
\hline \textbf{Step 10} 
& $ C_{12} = [11, 12]$, $ C_{6} = [1, 2, 3, 4, 5, 6]$, $ C_{8} = [8, 10, 13]$\\
\hline \textbf{Step 11} 
& $ C_{12} = [11, 12]$, $ C_{6} = [1, 2, 3, 4, 5, 6]$, $ C_{8} = [8, 10, 13]$\\
\hline \textbf{Step 12} 
& $ C_{12} = [11, 12]$, $ C_{6} = [1, 2, 3, 4, 5, 6]$, $ C_{8} = [8, 10]$, \\
& $ C_{9} = [9, 13]$\\
\hline \textbf{Step 13} 
& $ C_{12} = [11, 12]$, $ C_{6} = [1, 2, 3, 4, 5, 6]$, $ C_{8} = [8, 10]$, \\
& $ C_{9} = [9, 13]$\\
\hline \textbf{Step 14} 
& $ C_{6} = [1, 2, 3, 4, 5, 6]$, $ C_{8} = [8, 10]$, $ C_{9} = [9, 11, 13]$\\
\hline \textbf{Step 15} 
& $ C_{6} = [1, 2, 3, 4, 5, 6]$, $ C_{8} = [8, 10]$, $ C_{9} = [9, 11, 13]$\\
\hline \textbf{Step 16} 
& $ C_{6} = [1, 2, 3, 4, 5, 6]$, $ C_{8} = [8, 10]$ \\
& $ C_{9} = [9, 11, 12, 13]$\\
\hline \textbf{Step 17} 
& $ C_{6} = [1, 2, 3, 4, 5, 6]$, $ C_{9} = [8, 9, 11, 12, 13]$\\
\hline \textbf{Step 18} 
& $ C_{6} = [1, 2, 3, 4, 5, 6]$, $ C_{9} = [8, 9, 11, 12, 13]$\\
\hline \textbf{Step 19} 
& $ C_{6} = [1, 2, 3, 4, 5, 6]$, $ C_{9} = [8, 9, 11, 12, 13]$\\
\hline \textbf{Step 20} 
& $ C_{6} = [1, 2, 3, 4, 5, 6]$, $ C_{9} = [7, 8, 9, 11, 12, 13]$\\
\hline \textbf{Steps 21} $\rightarrow$ \textbf{22}
& $ C_{6} = [1, 2, 3, 4, 5, 6, 7]$, $ C_{9} = [8, 9, 11, 12, 13]$\\
\hline \textbf{Steps 23} $\rightarrow$ \textbf{25}
& $ C_{6} = [1, 2, 3, 4, 5, 6, 7]$, $ C_{9} = [8, 9, 10, 11, 12, 13]$\\
\hline
\end{tabular}}
\caption{\label{fig-example}Example of SCoDA execution on a small network:
\textmd{
The order of arrival of the edges is indicated
as a label on each edge.
The table lists the communities with more than two nodes
for each step of the algorithm.
It shows the execution of SCoDA for $D=4$.
}}
\end{figure}

An example of execution of the algorithm on a toy network of 13 nodes
and 25 edges is shown in Figure \ref{fig-example}. Observe that this
execution of SCoDA is able to perfectly recover the two underlying
communities. This depends on the random shuffling of the edges, however,
and another instance may give a different output. In the next section,
we analyse the ability of SCoDA to detect communities in real-life
graphs, as well as the variance of the results provided by different
executions of the algorithm.

Note that when an edge \((u,v)\) arrives, only the community memberships
of \(u\) and \(v\) are modified. In particular there is \emph{no
propagation} of a community change to the neighbors of \(u\) or \(v\).
In Figure \ref{fig-example}, we see for instance that, when edge \(20\)
is streamed, only node \(7\) is transferred to community \(C_9\). A
consequence of this absence of propagation is that SCoDA is
\emph{embarrassingly parallel}: the execution of the algorithm can be
split into tasks, each processing a subset of the
edges\footnote{The choice of these subsets does not
matter since edges are considered in
random order.}, with \(d\) and \(c\) stored in a shared memory.

We also remark that the algorithm easily extends to weighted graph.
Indeed, we can consider that each edge is drawn with a probability
proportional to its weight instead of considering uniform probabilities.

\subsection{Complexity}\label{complexity}

The algorithm contains three parts: the initialization of the vectors
\(d\) and \(c\), which is linear in \(n\), the shuffling of the list of
edges, which is linear in \(m\) with the Fisher-Yates algorithm
\citep{fisher1938statistical}, and the main loop which is also linear in
\(m\). Thus, the time complexity of the algorithm is linear in \(m\)
(assuming that \(m\) is larger than \(n\), which is the case in
practice).

Concerning the space complexity, we only use two arrays of integers of
size \(n\), \(d\) and \(c\). Note that the algorithm does not need to
store the list of edges in memory, but can simply read it in a random
order, which is the main benefit of the streaming approach. Hence, the
space complexity of the algorithm is
\(2n \cdot \texttt{sizeOf(int)} = O(n)\).

\subsection{Degree threshold}\label{degree-threshold}

In the rest of the paper, the only parameter of SCoDA, \(D\), is set to
the \emph{mode} of the degree distribution of the network, i.e., the
degree that appears most often in the graph, excluding the leaf nodes.
Hence we take \(D= d_{\text{mode}}\) with

\begin{equation}\label{eq:mode}
d_{\text{mode}} =\arg \max_{d > 1} |\{ u\in V: d(u) = d \}|.
\end{equation}

This choice is justified in Section \ref{section-parameter}.

Note that the computation of \(d_{\text{mode}}\) for a given graph is
linear in the number of edges \(m\) and can be done in a streaming way
like SCoDA (but before the execution of SCoDA). Indeed, it is sufficient
to know the degree of each node; computing \(d_{\text{mode}}\) then
requires \(n\) comparisons.

\section{Experimental results}\label{experimental-results}

\label{section-experiments}

\subsection{Datasets}\label{datasets}

We use real-life networks provided by the Stanford Social Network
Analysis Project (SNAP \citep{yang2015defining}) for the experimental
evaluation of SCoDA. These datasets include ground-truth community
memberships that we use to measure the quality of the detection. We
consider datasets of different natures:

\begin{itemize}
\tightlist
\item
  \textbf{Social networks}: The YouTube, LiveJournal, Orkut and
  Friendster datasets correspond to social networks
  \citep{backstrom2006group}\citep{mislove2007measurement} where nodes
  represent users and edges connect users who have a friendship
  relation. In all these networks, users can create groups that are used
  as ground-truth communities in the dataset definitions.
\item
  \textbf{Co-purchasing network}: The Amazon dataset corresponds to a
  product co-purchasing network \citep{leskovec2007dynamics}. The nodes
  of the graph represent Amazon products and the edges correspond to
  frequently co-purchased products. The ground-truth communities are
  defined as the product categories.
\item
  \textbf{Co-citation network}: The DBLP dataset corresponds to a
  scientific collaboration network \citep{backstrom2006group}. The nodes
  of the graph represent the authors and the edges the co-authorship
  relations. The scientific conferences are used as ground-truth
  communities.
\end{itemize}

The size of these networks ranges from approximately one million edges
to more than one billion edges. It enables us to test the ability of
SCoDA to scale to very large networks. The characteristics of these
datasets can be found in Table \ref{snap-datasets}.

\begin{table}[h]
\begin{tabular}{l | r r r}
& $|V|$ & $|E|$ & \# communities\\
\hline
Amazon & 334,863 & 925,872 & 311,782  \\
DBLP & 317,080 & 1,049,866 & 1,449,666 \\
YouTube & 1,134,890 & 2,987,624 & 8,455,253 \\
LiveJournal & 3,997,962 & 3,4681,189 & 137,177 \\
Orkut & 3,072,441 & 117,185,083 & 49,732 \\
Friendster & 65,608,366 & 1,806,067,135 & 2,547 \\
\end{tabular}
\caption{SNAP datasets used for the benchmark on real networks}
\label{snap-datasets}
\end{table}

\subsection{Benchmark algorithms}\label{benchmark-algorithms}

For assessing the performance of SCoDA we use a wide range of
state-of-the-art algorithms that are based on various approaches:

\begin{itemize}
\tightlist
\item
  \textbf{SCD} (S) partitions the graph by maximizing the WCC, which is
  a community quality metric based on triangle counting
  \citep{prat2014high}.
\item
  \textbf{Louvain} (L) is based on the optimization of the well-known
  modularity metric \citep{blondel2008fast}.
\item
  \textbf{Infomap} (I) splits the network into modules by compressing
  the information flow generated by random walks
  \citep{rosvall2008maps}.
\item
  \textbf{Walktrap} (W) uses random walks to estimate the similarity
  between nodes, which is then used to cluster the network
  \citep{pons2005computing}.
\item
  \textbf{OSLOM} (O) partitions the network by locally optimizing a
  fitness function which measures the statistical significance of a
  community \citep{lancichinetti2011finding}.
\end{itemize}

\subsection{Performance metrics}\label{performance-metrics}

We use two metrics for the performance evaluation of the selected
algorithms. The first is the \emph{average F1-score}
\citep{yang2013overlapping}\citep{prat2014high}. Given an estimate
\(\hat C\) of a true community \(C\), the precision and recall of this
estimation, that respectively penalize false positive and false
negative, are defined as:
\[\textrm{Precision}(\hat C, C) = \frac{|\hat C \cap C|}{|\hat C|}
,\ \textrm{Recall}(\hat C, C) = \frac{|\hat C \cap C|}{|C|}.\] The
F1-Score of the estimation \(\hat C\) of \(C\) is then defined as the
harmonic mean of precision and recall:
\[\text{F1}(\hat C,C) = 2 \frac{\textrm{Precision}(\hat C,C) \cdot
\textrm{Recall}(\hat C,C)}{\textrm{Precision}(\hat C,C) +
\textrm{Recall}(\hat C,C)}.\] Now consider some partition of the graph
into \(K\) communities, \(\mathcal{C} = \{ C_1,...,C_K \}\). The
F1-Score of the partition
\(\hat{\mathcal{C}} = \{ \hat C_1,...,\hat C_L \}\) with respect to
\(\mathcal{C}\) is defined by:
\[ \text{F1}(\hat{\mathcal{C}}, \mathcal{C}) =
\frac{1}{K} \sum_{k=1}^K \max_{1 \leq l \leq L} \text{F1}(\hat C_l, C_k).\]
Finally, the average F1-Score between the set of detected communities
\(\hat{\mathcal{C}}\) and the set of ground-truth communities
\(\mathcal{C}\) is:
\[\overline{\text{F1}}(\hat{\mathcal{C}}, \mathcal{C})
= \frac{\text{F1} (\hat{\mathcal{C}},\mathcal{C}) + \text{F1} (\mathcal{C},\hat{\mathcal{C}})}{2}\]

The second metric we use is the \emph{Normalized Mutual Information}
(NMI), which is based on the mutual entropy between indicator functions
for the communities \citep{lancichinetti2009detecting}.

\subsection{Benchmark setup}\label{benchmark-setup}

The experiments were performed on EC2 instances provided by Amazon Web
Services of type \texttt{m4.4xlarge} with 64 GB of RAM, 100 GB of disk
space, 16 virtual CPU with Intel Xeon Broadwell or Haswell and Ubuntu
Linux 14.04 LTS.

SCoDA is implemented in C++ and the source code can be found on
GitHub\footnote{\url{https://github.com/ahollocou/scoda}}. For the other
algorithms, we used the C++ implementations provided by the authors,
that can be found on their respective websites. Finally, all the scoring
functions were implemented in C++. We used the implementation provided
by the authors of \citep{lancichinetti2009detecting} for the NMI and the
implementation provided by the authors of SCD \citep{prat2014high} for
the F1-Score.

\subsection{Benchmark results}\label{benchmark-results}

\subsubsection*{Execution time}\label{execution-time}
\addcontentsline{toc}{subsubsection}{Execution time}

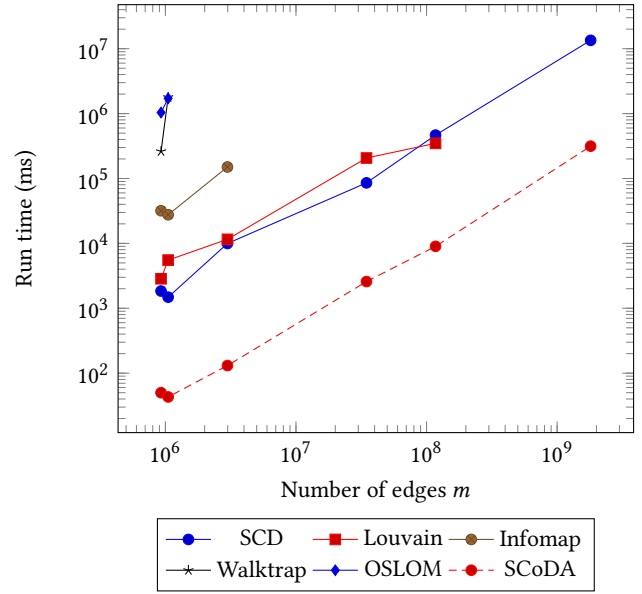
\begin{figure}[h]
\begin{center}
\begin{tikzpicture}
\begin{loglogaxis}[
legend style={at={(0.5,-0.20)},
anchor=north,legend columns=3},
xlabel=Number of edges $m$,
ylabel=Run time (ms)]
\addplot coordinates { (925872, 1843.0)(1049866, 1484.0)(2987624, 9964.0)(34681189, 85701.0)(117185083, 466262.0)(1806067135, 1.3463853e7) };
\addlegendentry{SCD}
\addplot coordinates { (925872, 2851.26)(1049866, 5523.61)(2987624, 11549.37)(34681189, 206872.54)(117185083, 348345.81) };
\addlegendentry{Louvain}
\addplot coordinates { (925872, 31842.75)(1049866, 27622.8)(2987624, 150469.58) };
\addlegendentry{Infomap}
\addplot coordinates { (925872, 261421.21)(1049866, 1.78522095e6) };
\addlegendentry{Walktrap}
\addplot coordinates { (925872, 1.03853515e6)(1049866, 1.71769557e6) };
\addlegendentry{OSLOM}
\addplot coordinates { (925872, 50.0)(1049866, 43.0)(2987624, 131.0)(34681189, 2581.0)(117185083, 9001.0)(1806067135, 314923.0) };
\addlegendentry{SCoDA}
\end{loglogaxis}
\end{tikzpicture}
\end{center}
\caption{Run times in milliseconds with respect to the number of edges for SNAP networks}
\label{figure-snap-run-time}
\end{figure}

We compare the execution times of the different algorithms on SNAP
networks in Table \ref{table-snap-run-time}. The entries that are not
reported in the table corresponds to algorithms that returned execution
errors or algorithms with execution times exceeding \(6\) hours. In our
experiments, only SCD, except from SCoDA, was able to run on all
datasets. The fastest algorithms in our benchmarks are SCD and Louvain
and we observe that they run more than ten times slower than our
streaming algorithm. More precisely, SCoDA runs in less than 50ms on the
Amazon and DBLP networks, which contain millions of edges, and in 5
minutes on the largest network, Friendster, that has more than one
billion edges. In comparison, it takes seconds for SCD and Louvain to
detect communities on the smallest networks, and several hours to run on
Friendster. Figure \ref{figure-snap-run-time} shows the execution times
of all the algorithms with respect to the number of edges in the
network. We remark that there is more than one order of magnitude
between SCoDA and the other algorithms.

In order to compare the execution time of SCoDA with a minimal algorithm
that only reads the list of edges without doing any additional
operation, we measured the run time of the Unix command \texttt{cat} on
the largest dataset, Friendster. \texttt{cat} reads the edge file
sequentially and writes each line corresponding to an edge to standard
output. In our expermiments, the command \texttt{cat} takes 152 seconds
to read the list of edges of the Friendster dataset, whereas SCoDA
processes this network in 314 seconds. That is to say, reading the edge
stream is only twice faster than the execution of SCoDA.

\begin{table}[h]
\begin{center}
\begin{tabular}{l | r r r r r r r}
 & S & L & I & W & O & SCoDA \\
\hline
Amazon & 1.84 & 2.85 & 31.8 & 261 & 1038 & \textbf{0.04} \\
DBLP & 1.48 & 5.52 & 27.6 & 1785 & 1717 & \textbf{0.04} \\
YouTube & 9.96 & 11.5 & 150 & - & - & \textbf{0.13} \\
LiveJournal & 85.7 & 206 & - & - & - & \textbf{2.58} \\
Orkut & 466 & 348 & - & - & - &  \textbf{9.26} \\
Friendster & 13464 & - & - & - & -  & \textbf{314} \\
\end{tabular}
\end{center}
\caption{Execution times in seconds}
\label{table-snap-run-time}
\end{table}

\subsubsection*{Memory consumption}\label{memory-consumption}
\addcontentsline{toc}{subsubsection}{Memory consumption}

We measured the memory consumption of SCoDA and compared it to the
memory that is needed to store the list of the edges for each network,
which is a lower bound of the memory consumption of the other
algorithms. We use 64-bit integers to store the node indices. The memory
needed to represent the list of edges is 14,8 MB for the smallest
network, Amazon, and 28,9 GB for the largest one, Friendster. In
comparison, SCoDA consumes 5,4 MB on Amazon and only 1,1 GB on
Friendster.

\subsubsection*{Detection scores}\label{detection-scores}
\addcontentsline{toc}{subsubsection}{Detection scores}

Tables \ref{table-snap-f1-score} and \ref{table-snap-nmi} show the
Average F1-score and NMI of the algorithms on the SNAP datasets. Note
that the NMI on the Friendster dataset is not reported in the table
because the scoring program used for its computation
\citep{lancichinetti2009detecting} cannot handle the size of the output
on this dataset. While Louvain and OSLOM clearly outperform SCoDA on
Amazon and DBLP (at the expense of longer execution times), SCoDA shows
similar performance as SCD on YouTube and much better performance than
SCD and Louvain on LiveJournal, Orkut and Friendster (the other
algorithms do not run these datasets). Thus SCoDA does not only run much
faster than the existing algorithms but the quality of the detected
communities is also better than that of the state-of-the-art algorithms
for very large networks.

\begin{table}
\begin{center}
\begin{tabular}{l | r r r r r r}
& S & L & I & W & O & SCoDA \\
\hline
Amazon & 0.39 & \textbf{0.47} & 0.30 & 0.39 & \textbf{0.47} & 0.37 \\
DBLP & 0.30 & 0.32 & 0.10 & 0.22 & \textbf{0.35} & 0.23 \\
YouTube & 0.23 & 0.11 & 0.02 & - & - & \textbf{0.24} \\
LiveJournal & 0.19 & 0.08 & - & - & - & \textbf{0.27} \\
Orkut & 0.22 & 0.19 & - & - & - & \textbf{0.42} \\
Friendster & 0.10 & - & - & - & - & \textbf{0.20} \\
\end{tabular}
\end{center}
\caption{Average F1 Scores}
\label{table-snap-f1-score}
\end{table}

\begin{table}
\begin{center}
\begin{tabular}{l | r r r r r r}
& S & L & I & W & O & SCoDA \\
\hline
Amazon & 0.16 & 0.24 & 0.16 & \textbf{0.26} & 0.23 & 0.12 \\
DBLP & \textbf{0.15} & 0.14 & 0.01 & 0.10 & \textbf{0.15} & 0.03 \\
YouTube & \textbf{0.10} & 0.04 & 0.00 & - & - & \textbf{0.10} \\
LiveJournal & 0.05 & 0.02 & - & - & - & \textbf{0.09} \\
Orkut & 0.11 & 0.05 & - & - & - & \textbf{0.22} \\
\end{tabular}
\end{center}
\caption{NMI}
\label{table-snap-nmi}
\end{table}

\subsection{Variance of the algorithm}\label{variance-of-the-algorithm}

SCoDA is not a deterministic algorithm as it depends on a random
permutation of the list of the edges. We study its variance over
multiple runs by computing the standard deviation of the F1-Score and of
the number of communities over 1000 independent runs. The results are
collected in Table \ref{table-snap-variance}. We see that the standard
deviation is 100 times lower than the average value of these metrics.
Note that these experiments were only run on the smaller datasets
because the run times on bigger networks were prohibitive.

\begin{table}
\begin{tabular}{l | c | c | c}
 & Average F1-Score & Nb of communities \\
 \hline
 Amazon & $ 3.7\times 10^{-1} \pm 5 \times 10^{-4} $ & $ 1.0\times 10^{5} \pm 1 \times 10^{2} $ \\
 DBLP & $ 2.3\times 10^{-1} \pm 4\times 10^{-4} $ & $ 1.9\times 10^{5} \pm 2 \times 10^{2} $ \\
 YouTube & $ 2.4\times 10^{-1} \pm 6 \times 10^{-4} $ & $ 8.8\times 10^{5} \pm 2 \times 10^{2} $ \\
 LiveJournal & $ 2.7\times 10^{-1} \pm 2 \times 10^{-4} $ & $ 2.7\times 10^{6} \pm 6 \times 10^{2} $ \\
\end{tabular}
\caption{Variance analysis of SCoDA on SNAP (average value $\pm$ standard deviation)}
\label{table-snap-variance}
\end{table}

\section{Theoretical analysis}\label{theoretical-analysis}

\label{section-analysis}

In this section, we provide a theoretical analysis of SCoDA explaining
its good performance. We consider successively the precision and recall
of the algorithm.

\subsection{Notation}\label{notation}

Observe that, as far as the analysis is concerned, we can assume that
the algorithm performs at each step a random sampling without
replacement of the edges, instead of the initial random shuffling. When
an edge \(e=(u,v)\) is drawn, there are two cases depending on the
degrees \(d_u\) and \(d_v\).

\begin{itemize}
\tightlist
\item
  If \(d_u \leq D\) and \(d_v \leq D\) then \(u\) joins the community of
  \(v\) or conversely, \(v\) joins the community of \(u\). In this case
  we say that \(e\) is a \emph{transfer edge} and we use \(t(e)\) to
  denote this event.
\item
  Otherwise, \(u\) and \(v\) remain in their communities. We say that
  \(e\) is a \emph{blank edge}.
\end{itemize}

\subsection{Precision}\label{precision}

\label{subsection-precision}

In this part of the analysis, we are interested in the false positives
detected by SCoDA.

Let \(C \subset V\) be the community that we want to detect. Let \(S\)
be a community returned by SCoDA such that \(S \cap C \neq \emptyset\).
Note that there is necessarily such a \(S\) because the algorithm
performs a partition of the set of nodes \(V\). We say that node
\(u \in V\) is a \emph{false positive} in \(S\) with respect to
community \(C\) if \(u\in S \setminus C\). Note that if SCoDA returns a
false positive, then one of the edges between \(C\) and
\(\overline{C} = V \setminus C\) is necessarily a transfer edge. Here,
we study the quantity of such edges that can lead to false positives,
that we call \emph{false-positive edges}. We use \(\text{FPE}(C)\) to
denote their number:
\[\text{FPE}(C) = |{e \in e(C, \overline{C}): t(e)}|\]

Let \((u,v)\) be an edge of \(e(C,\overline{C})\). In what follows, we
consider by convention that, for such edges, \(u\in C\) and
\(v\in \overline{C}\). We observe that if the first \(D\) edges that are
drawn in \(e(\{u\})\) (i.e., edges with an end equal to \(u\)) are in
\(e(C,C)\) (intra-community edges), then \((u,v)\) cannot be a transfer
edge because \(d_u > D\) when \((u,v)\) is streamed. The probability for
the \(k^{\text{th}}\) edge in \(e({u})\) to be in \(e(C,C)\) knowing
that the previous edges were in \(e(C,C)\) is:
\[1 - \frac{d_{\overline{C}}(u)}{d(u) - k}\]

Therefore, the expected value of \(\text{FPE}(C)\) satisfies:
\[\begin{split}
\mathbb{E}[\text{FPE}(C)] &= \sum_{e\in e(C,\overline{C})} \mathbb{P}[t(e)] \\
& \leq
\sum_{(u,v)\in e(C,\overline{C})}
\left[ 1 - \prod_{k=0}^{D - 1} \left(1 - \frac{d_{\overline{C}}(u)}{d(u) - k}\right) \right]
\end{split}
\] The quantity \(d_{\overline{C}}(u)/d(u)\) is known as the \emph{Out
Degree Fraction} (ODF) of node \(u\) for the community \(C\). We use
\(\text{ODF}(u, C)\) to denote this quantity. Observe that a \emph{good}
community corresponds to low values of \(\text{ODF}(u, C)\) for
\(u \in C\).

The ratio of the expected number of false positive edges to the total
number of edges in the community satisfies:
\[ \frac{\mathbb{E}[\text{FPE}(C)]}{|e(C)|}
\leq \phi_0(C) \left[ 1 -
\frac{
\sum_{(u,v)\in e(C,\overline{C})}
\prod_{k=0}^{D - 1} \left(1 - \frac{\text{ODF}(u,C)}{1 - \frac{k}{d(u)}} \right)
}{|e(C, \overline{C})|}\right],\] where \(\phi_0(C)\) is the
pseudo-conductance defined in Section \ref{section-algorithm}. Note that
the term
\(\prod_{k=0}^{D - 1} \left(1 - \frac{\text{ODF}(u,C)}{1 - k/d(u)} \right)\)
is null if \(D > d_C(u)\). We remark that the lower the
pseudo-conductance \(\phi_0(C)\) and the Out Degree Fraction
\(\text{ODF}(u, C)\) for \(u\in \partial C\), the less false-positive
edges are observed. For \emph{good} communities \(C\), these quantities
are typically small, leading to a good precision.

\subsection{Recall}\label{recall}

Now we analyze the performance of SCoDA in terms of recall, i.e., its
ability to recover all the nodes of a given community \(C\).

\subsubsection*{Intuition}\label{intuition}
\addcontentsline{toc}{subsubsection}{Intuition}

We would like to have a low probability of splitting a community \(C\)
into several sub-communities. We remark that SCoDA may split a community
\(C\) into two sub-communitites \(C_1\) and \(C_2\) if the edges
\((u,v) \in e(C_1,C_2)\) (\(u\in C_1\) and \(v \in C_2\)) satisfy
\(d_{C_1}(u) > D\) or \(d_{C_2}(v) > D\). In this case, when
\((u,v) \in e(C_1, C_2)\) is processed, we can potentially have
\(d_u > D\) or \(d_v > D\).

Now if the community \(C\) is homogeneous, we have typically
\(d_{C_1}(u) \simeq \bar{d} \frac{|C_1|}{|C|}\) and
\(d_{C_2}(u) \simeq \bar{d} \frac{|C_2|}{|C|}\), where \(\bar{d}\)
corresponds to the average intra-community degree in \(C\). Therefore,
our community \(C\) is potentially split into \(C_1\) and \(C_2\) if:
\[ \bar{d} \frac{|C_1|}{|C|} > D \text{ or } 
\bar{d} \frac{|C_2|}{|C|} > D \] In particular, if the parameter \(D\)
is close or larger than \(\bar{d}\), then these inequalities are not
satisfied and the probability of splitting \(C\) into sub-clusters is
low.

\subsubsection*{Analysis on random
graphs}\label{analysis-on-random-graphs}
\addcontentsline{toc}{subsubsection}{Analysis on random graphs}

The previous argument is not rigorous and only provides an insight into
the behavior of SCoDA. In order to study the ability of SCoDA to recover
an entire community, we study its results on a random graph model. We
represent a community as a small, homogeneous and well-connected random
graph. For such a random graph, we expect SCoDA to return the entire
graph as a community. Hence, we measure the average value of the
relative size of the largest community returned by the algorithm:
\[\max_{\hat{C} \in \hat{\mathcal{C}}} \frac{|\hat{C}|}{n} \]

We choose the simplest random graph model, the \text{Erd\H{o}s R\'enyi}
graph \citep{gilbert1959random} (other experiments, not reported here,
showed the same type of results on the configuration model
\citep{molloy1995critical}\citep{newman2001random}). Recall that this
model has two parameters \(n\) and \(p\): \(n\) is the number of nodes,
and each edge \((u,v)\) is included in the graph with probability \(p\)
independently of every other edge. Although this model is inappropriate
for modelling entire real-life networks, it is reasonable for
representing small communities.

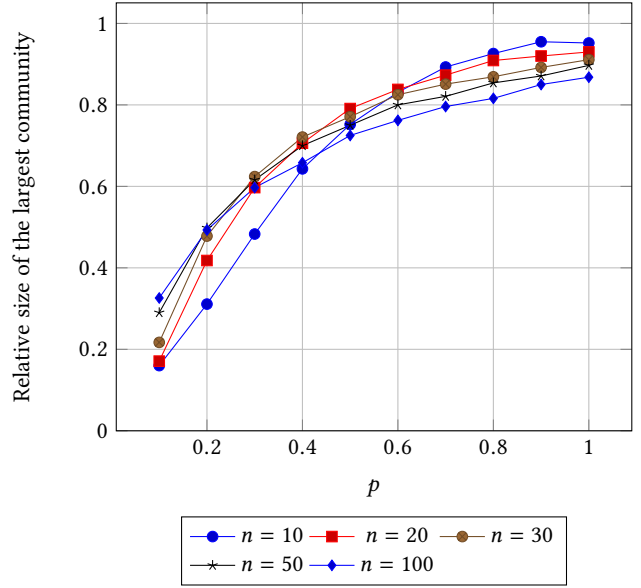
\begin{figure}
\begin{center}
\begin{tikzpicture}
\begin{axis}[
grid=major,
ylabel=Relative size of the largest community,
xlabel=$p$,
ymin=0,
legend style={at={(0.5,-0.20)},
anchor=north,legend columns=3},
]
\addlegendentry{$n=10$}
\addplot coordinates {(0.1, 0.16) (0.2, 0.311) (0.3, 0.483) (0.4, 0.643) (0.5, 0.752) (0.6, 0.831) (0.7, 0.893) (0.8, 0.926) (0.9, 0.955) (1.0, 0.952) };
\addlegendentry{$n=20$}
\addplot coordinates {(0.1, 0.171) (0.2, 0.418) (0.3, 0.597) (0.4, 0.706) (0.5, 0.791) (0.6, 0.838) (0.7, 0.873) (0.8, 0.909) (0.9, 0.92) (1.0, 0.93) };
\addlegendentry{$n=30$}
\addplot coordinates {(0.1, 0.217) (0.2, 0.478) (0.3, 0.624) (0.4, 0.721) (0.5, 0.771) (0.6, 0.825) (0.7, 0.851) (0.8, 0.869) (0.9, 0.892) (1.0, 0.911) };
\addlegendentry{$n=50$}
\addplot coordinates {(0.1, 0.29) (0.2, 0.498) (0.3, 0.616) (0.4, 0.7) (0.5, 0.75) (0.6, 0.8) (0.7, 0.821) (0.8, 0.854) (0.9, 0.871) (1.0, 0.897) };
\addlegendentry{$n=100$}
\addplot coordinates {(0.1, 0.326) (0.2, 0.493) (0.3, 0.597) (0.4, 0.658) (0.5, 0.725) (0.6, 0.762) (0.7, 0.796) (0.8, 0.816) (0.9, 0.85) (1.0, 0.868) };
\end{axis}
\end{tikzpicture}
\caption{Average relative size of the largest community returned by SCoDA
on a Erd\H{o}s-R\'enyi random graph of parameters $(n,p)$}
\label{figure-er}
\end{center}
\end{figure}

We experimentally generate graphs using this model for different values
of \(n\) and \(p\). We plot the average value of the relative size of
the largest community returned by the algorithm in Figure
\ref{figure-er}. For each value of \((n, p)\), \(1000\) graphs were
generated and SCoDA was run once on each graph. The degree threshold
\(D\) was set to \(d_{\text{mode}}\) as in previous experiments.

We see that the relative size of the largest community is close to \(1\)
when the parameter \(p\) tends to \(1\). The case \(p=1\) corresponds to
the situation where the graph is complete and where we want to recover
the densest possible community, the clique. We see that SCoDA is able to
recover almost the entire community in this situation. Besides, as soon
as \(p>0.5\), SCoDA recovers more than \(75\%\) of the nodes in its
largest community.

\section{Setting the degree
threshold}\label{setting-the-degree-threshold}

\label{section-parameter}

In this section, we give a deeper insight into the choice of the only
parameter of SCoDA, the degree threshold \(D\).

\subsection{Some strategies}\label{some-strategies}

Consider the dynamic graph \(H\) having the same nodes as \(G\) and
whose edges are those successively considered by SCoDA. Note that \(D\)
corresponds to the maximum updated degree in this graph until a change
in the communities occurs. Indeed, when an edge \((u,v)\) is streamed,
there is a change in the communitites if and only if both \(d_u\) and
\(d_v\) are lower than \(D\).

As observed before, with high probability, \emph{intra-community} edges
are streamed before \emph{inter-community} edges. If \(D\) is too low,
SCoDA might not take some \emph{intra-community} edges into account and
can potentially split communities into sub-clusters. Remark that, if
\(D=1\), then the algorithm only outputs communities with at most two
nodes. On the contrary, if \(D\) is too high, community transfers due to
\emph{inter-community} edges will occur and deteriorate the quality of
the detected communities.

Hence, \(D\) is intuitively related to the degree distribution of the
graph \(G(V,E)\). We can consider several options for the choice of this
parameter:

\begin{itemize}
\tightlist
\item
  \textbf{Average degree}: \(D=d_{\text{avg}}\), the average value of
  the degree \(d(u)\) over the network;
\item
  \textbf{Median degree}: \(D=d_{\text{med}}\), the median value of the
  degree \(d(u)\) over the network;
\item
  \textbf{Mode of the degree distribution}: \(D=d_{\text{mode}}\), the
  most common degree in the network excluding leaf nodes, as defined by
  \eqref{eq:mode}.
\end{itemize}

Recall that we have chosen \(D=d_{\text{mode}}\) until now. We justify
this choice below both experimentally on SNAP datasets and theoretically
using the previous analysis.

\subsection{Experimental analysis}\label{experimental-analysis}

We evaluate the performance of SCoDA as a function of \(D\) on real-life
networks. For this purpose, we perform experiments on the SNAP datasets
introduced above and we use the Average F1 Score to measure the quality
of the detected communities. In order to evaluate the accuracy of the
different choices of \(D\), we consider the \emph{relative quality
ratio} \(Q\) between the F1-score obtained for a given \(D\), noted
\(\overline{\text{F1}}(D)\), and the maximum of this score observed for
any value of \(D\):
\[ Q(D) = \frac{\overline{\text{F1}}(D)}{\max_{D'} \overline{\text{F1}}(D')} \]

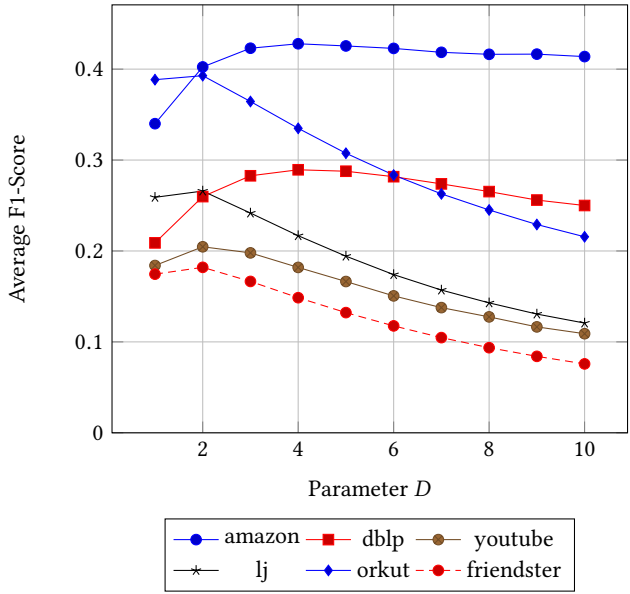
\begin{figure}
\begin{center}
\begin{tikzpicture}
\begin{axis}[
xlabel=Parameter $D$,
ylabel=Average F1-Score,
grid=major,
ymin=0,
legend style={at={(0.5,-0.20)},
anchor=north,legend columns=3},
]
\addplot coordinates {
(1, 0.34) (2, 0.4024) (3, 0.423) (4, 0.4279) (5, 0.4255) (6, 0.4228) (7, 0.4185) (8, 0.4163) (9, 0.4165) (10, 0.4138)
};
\addlegendentry{amazon}
\addplot coordinates {
(1, 0.2089) (2, 0.2598) (3, 0.2827) (4, 0.2893) (5, 0.2877) (6, 0.2817) (7, 0.2738) (8, 0.2653) (9, 0.256) (10, 0.25)
};
\addlegendentry{dblp}
\addplot coordinates {
(1, 0.184) (2, 0.2046) (3, 0.1979) (4, 0.1819) (5, 0.1664) (6, 0.1506) (7, 0.1377) (8, 0.1275) (9, 0.1165) (10, 0.109)
};
\addlegendentry{youtube}
\addplot coordinates {
(1, 0.259) (2, 0.2659) (3, 0.2417) (4, 0.2168) (5, 0.1943) (6, 0.174) (7, 0.157) (8, 0.1429) (9, 0.1306) (10, 0.1207)
};
\addlegendentry{lj}
\addplot coordinates {
(1, 0.3884) (2, 0.3928) (3, 0.3644) (4, 0.3349) (5, 0.3075) (6, 0.2835) (7, 0.2628) (8, 0.2451) (9, 0.2291) (10, 0.2156)
};
\addlegendentry{orkut}
\addplot coordinates {
(1, 0.1745) (2, 0.182) (3, 0.1665) (4, 0.1486) (5, 0.1322) (6, 0.1176) (7, 0.1047) (8, 0.0936) (9, 0.0841) (10, 0.0758)
};
\addlegendentry{friendster}
\end{axis}
\end{tikzpicture}
\caption{Average F1 Score with respect to $D$ for SNAP networks}
\label{figure-snap-d}
\end{center}
\end{figure}

In Figure \ref{figure-snap-d}, we plot the Average F1-Score of SCoDA
with respect to \(D\). Note that, the singletons returned by SCoDA were
excluded from the computation of the F1-Score in order to boost the
execution time of this scoring metric for different values of \(D\),
which explains why the scores slightly differ from the ones listed in
Table \ref{table-snap-f1-score}. As expected, we observe that the
F1-Score first increases with \(D\) until it reaches a maximum value,
and then decreases as \(D\) continues to increase.

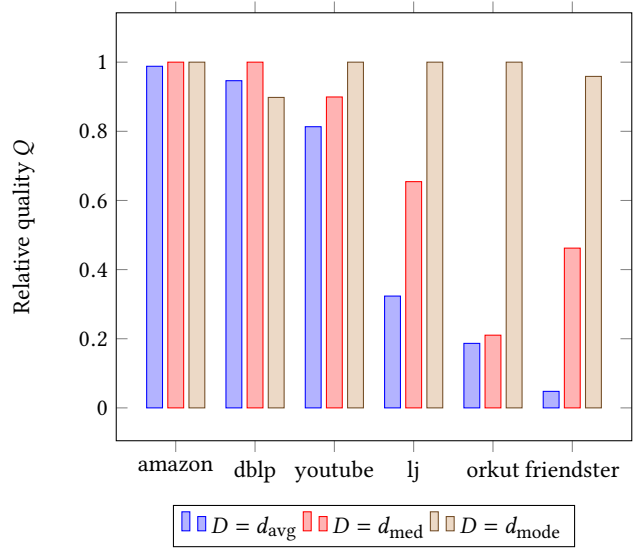
\begin{figure}
\begin{tikzpicture}
\begin{axis}[
symbolic x coords={amazon,dblp,youtube,lj,orkut,friendster},
ylabel=Relative quality $Q$,
enlargelimits=0.15,
legend style={at={(0.5,-0.15)},
anchor=north,legend columns=-1},
ybar,
bar width=6pt,
]
\addplot
coordinates {(amazon, 0.9881) (dblp, 0.9464) (youtube, 0.8133) (lj, 0.3234) (orkut, 0.1866) (friendster, 0.0478) };
\addplot
coordinates {(amazon, 1.0) (dblp, 1.0) (youtube, 0.8993) (lj, 0.6544) (orkut, 0.2103) (friendster, 0.4621) };
\addplot
coordinates {(amazon, 1.0) (dblp, 0.898) (youtube, 1.0) (lj, 1.0) (orkut, 1.0) (friendster, 0.9588) };
\legend{$D=d_{\text{avg}}$,$D=d_{\text{med}}$,$D=d_{\text{mode}}$}
\end{axis}
\end{tikzpicture}
\caption{Relative quality $Q$ for different choices of $D$ on SNAP networks}
\label{figure-snap-q}
\end{figure}

In Figure \ref{figure-snap-q}, we plot the ratio \(Q(D)\) defined above
for the three choices for \(D\) listed above (average, median and mode).
We see that \(Q(D) > 0.9\) for \(D = d_{\text{mode}}\) whereas \(Q(D)\)
shows poor values for the other choices of \(D\) on the social networks
datasets LiveJournal, Orkut and Friendster. This justifies
experimentally our previous choice.

Table \ref{table-snap-degrees} collects different statistics on the
degree distribution of the datasets, including the values of
\(d_{\text{avg}}\), \(d_{\text{med}}\) and \(d_{\text{mode}}\). We
remark that for YouTube, LiveJournal, Orkut and Friendster the density
(\(\frac{m}{2n(n-1)}\)) is lower and the maximum degree
\(d_{\text{max}}\) is higher than for the DBLP and Amazon datasets. The
different nature of the networks could explain the differences in these
statistics and in the behavior of SCoDA. On the one hand, we have social
networks where the performance of SCoDA decreases drastically when \(D\)
increases, and, on the other hand, we have a co-citation and a
co-purchasing networks for which the F1-Score decreases much more
slowly.

Note that one could ask why we do not simply use a fixed value for D
(e.g. \(D=2\)) but experiments on the random graphs defined in
\S\ref{subsection-precision} shows that it deteriorates the recall of
the algorithm.

\begin{table}
\begin{tabular}{l | r r r r r}
& $d_{\text{avg}}$ & $ d_{\text{max}} $ & $ d_{\text{med}} $ & $ d_{\text{mode}} $ & density\\
\hline
Amazon & 5.5  & 549 & 4 & 4& 1.6e-5\\
DBLP & 6.6 & 343 & 4 & 2 & 2.1e-5\\
YouTube & 5.3 & 28754 & 1 & 2 & 4.6e-6\\
LiveJournal & 17.3 & 14815 & 6 & 2 & 4.3e-6\\
Orkut & 76.2 & 33313 & 45 & 2 & 2.5e-5\\
Friendster & 55.1 & 5214 & 9 & 2 & 8.4e-7\\
\end{tabular}
\caption{Statistics on the degree distribution of SNAP networks}
\label{table-snap-degrees}
\end{table}

\subsection{Theoretical insights}\label{theoretical-insights}

In Section \ref{section-analysis}, we observed that the ODFs of nodes at
the boundary of community \(C\) need to be higher than \(D\) in order to
obtain few false positive edges. Besides, we have seen that \(D\) must
be close to the intra-community degree in \(C\) in order to decrease the
likelihood for \(C\) to be split into sub-clusters by SCoDA. These
arguments suggest that the parameter \(D\) should be close to the
typical value of intra-community degrees in the network. Since most
nodes have few inter-community links, this is well approximated by the
most probable degree in the graph, that is the mode of the degree
distribution as chosen in our experiments. Observe that the average and
the median are not \emph{typical} values of the degree of a node, which
may explain the worse performance of SCoDA in these cases.

\section{Conclusion and future work}\label{conclusion-and-future-work}

\label{section-conclusion}

We introduced a novel approach for community detection based on a random
stream of edges. This approach is based on simple properties of such
edge streams, that are closely related to important concepts in network
analysis such as conductance and out-degree fraction. We designed an
algorithm, named SCoDA, that stores only two integers for each node and
runs linearly in the number of edges. In our experiments, SCoDA runs
more than 10 times faster than state-of-the-art algorithms such as
Louvain and SCD and shows better detection scores on the largest
networks. Thus SCoDA would be useful in many applications where massive
graphs arise. For example, the web graph contains around \(10^{10}\)
nodes which is much more than in the Friendster dataset.

While we evaluated the performance of the algorithm on static graphs
only, it would be interesting for future work to measure the ability of
SCoDA to handle evolving networks by conducting benchmarks on dynamic
datasets \citep{panzarasa2009patterns} with existing approaches
\citep{gauvin2014detecting}\citep{epasto2015efficient}. Note that
modifications to the algorithm design could be made to handle events
such as edge deletions.

Another interesting research direction would be to exploit the fact
that, between two runs of SCoDA, the \emph{transfer-edges} and the
\emph{blank-edges} can change. For each edge of the network, we could
count how many times it corresponds to a \emph{transfer-edge} over
several runs and use this result to refine the community detection with,
for instance, a boostrap aggregating approach
\citep{breiman1996bagging}.

Furthermore, we remark that the condition \(d_u \leq D\) and
\(d_v \leq D\) plays an important role in the definition of the
algorithm. Future works could explore different ways to define when an
edge arrives \emph{early} or \emph{late}. For instance, the general
condition \(f(d_u,d_v) \leq D\), could lead to different results for
certain choices of \(f\).

Finally, SCoDA only returns disjoint communities, whereas, in many real
networks, overlaps between communities can be observed
\citep{lancichinetti2009detecting}. An important research direction
would consist in adapting SCoDA to overlapping community detection and
compare it to existing approaches
\citep{xie2013overlapping}\citep{yang2013overlapping}.

\bibliographystyle{ACM-Reference-Format}

\end{document}